\documentclass[12pt]{article}
\usepackage{xcolor}
\usepackage{graphicx}
\usepackage{amsmath}
\usepackage{amssymb}
\usepackage{geometry}
\usepackage{array, tabularx}
\usepackage{multirow}
\usepackage{pdflscape}
\usepackage{amsfonts}
\usepackage{placeins}
\usepackage{hyperref}
\usepackage{url}
\usepackage{caption}
\usepackage{epstopdf}
\usepackage[normalem]{ulem}
\usepackage{subcaption}
\begin{document}
\title{Simple mathematical models for controlling COVID-19 transmission through social distancing and community awareness}

\author{Ahmed S. Elgazzar\\
\textit{Mathematics Department, Faculty of Science, Arish University}\\
\textit{45516 Arish, Egypt}\\
\url{aselgazzar@aru.edu.eg}}
\date{}
\maketitle

\begin{abstract}
The novel COVID-19 pandemic is a current, major global health threat. Up till now, there is no fully approved pharmacological treatment or a vaccine. In this study, simple mathematical models were employed to examine the dynamics of transmission and control of COVID-19 taking into consideration social distancing and community awareness. Both situations of homogeneous and nonhomogeneous population were considered. Based on the calculations, a sufficient degree of social distancing based on its reproductive ratio is found to be effective in controlling COVID-19, even in the absence of a vaccine. With a vaccine, social distancing minimizes the sufficient vaccination rate to control the disease. Community awareness also has a great impact in eradicating the virus transmission. The model is simulated on small-world networks and the role of social distancing in controlling the infection is explained.
\end{abstract}

\section{Introduction}
There is a growing interest in the study of mathematical epidemiology using a wide variety of mathematical models \cite{1}-\cite{4}. Epidemic models explains well the dynamics of both the transmission and the control of an infection. Numerous modeling approaches can be integrated to form a clearer picture of an infection and to establish measures on how to eradicate it. Epidemic models can be greatly improved when combined with real data.

The COVID-19 pandemic that reportedly originated in Wuhan, China, in December 2019 \cite{5} has now become a major global health threat, as well as a focus of interest in many researchers and the entire global community \cite{5}-\cite{11}. COVID-19 is found to be easily transmitted via close contact with an infective person, even during the incubation period \cite{8}. The basic reproductive ratio ($R_{0}$) is the average number of secondary infections arising from one infective person; it is a measure of the infectiousness of a disease. A recent study \cite{9} showed that COVID-19 has an $R_{0}$ of 3.28 in average; however, this has to be confirmed in further studies due to the novelty of the virus. Additionally, this pandemic has brought the concepts that community behavior has a primary role in the transmission or containment of the virus, and even the competing organizations have to collaborate among each other to contain the crisis \cite{10,11}.

Developing a COVID-19 vaccine is a very costly and time consuming process, where its effectiveness and safety have to be thoroughly ensured before public use. In some cases, vaccine-induced immunity wanes with time. Also, some viruses learn to mute and evade the immunity induced by vaccines. In the case of containing COVID-19, the most suitable strategy, for now, is to reduce its infectiousness by limiting contact with infective people and polluted surfaces in what is contemporarily popular as "social distancing (SD)" including the quarantine of infected individuals. The degree of SD can be regarded as the fraction of the entire population complying with the policies. Another interpretation considers the gradation of SD policies, i.e., closing borders to international travel; partially closing social venues, large malls, schools and universities, and prayer places; or total lockdown. The awareness of the community in applying these policies will have a great impact on the viral transmission control.

Adaptive behavior is extensively seen in multiagent biological systems \cite{12}, where the behavior of agents is coupled by their environment interaction. Such a coupling can be modeled by the Sato-Crutchfield formulation \cite{13,14} which corresponds to memory decay.  This formulation can also be equate to personal mistakes or environmental effects based on the community awareness.

Different mathematical modeling approaches for COVID-19  have been proposed \cite{15}-\cite{17}. A compartmental model is used in the present study, because of its simplicity and its ability to study many features of the dynamics of disease transmission and control. It should be noted that studying the equilibria of the associated dynamic system and its stability is of crucial importance as it helps in collecting information around the infection dynamics. As immunity to reinfection from COVID-19 lasts in a short time \cite{18} and, the virus can mute and evade the immunity, the SIRS model is suitable to examine the spread of COVID-19.

The significant role of the spatial aspect of the dynamics of infection transmission is emphasised by simulating epidemic models on networks \cite{19}. Regular lattices can display the clustering property only, while random graphs can show the long-range interactions. Small-world networks \cite{20} is one of the most important structures that interpolates between regular lattices and random graphs. Thus, both local and non-local interactions are combined. Small-world networks are used for studying many features of epidemics \cite{3,21}.

This study aims to develop simple compartmental models for COVID-19, which will be used to investigate the important roles of SD and community awareness in controlling the infection transmission. In Section \ref{s2}, a discussion of the SIRS model for the spread of COVID-19, with notes on the equilibria and the corresponding stability of the dynamic system is covered. The roles of SD, community awareness, and vaccination are explained, with provided calculations on the sufficient degree of SD and vaccination ratio to control COVID-19 transmission. Then, the model is simulated on small-world networks. In Section \ref{s3}, a nonhomogeneous population model is introduced; and the nonlocal interaction is highlighted. Some conclusions are summed up in Section \ref{s4}.

\section{Homogeneous population}\label{s2}
Assume that a homogenously mixed population of constant sufficiently large size can be divided into three compartments: susceptible ($ S $), infective (symptomatic or asymptomatic) ($ I $), and removed (isolated or recovered) ($ R $). Infection is transmitted via direct contact between susceptible and infective individuals at a constant rate $ \beta $. Infective individuals are removed at a constant rate $ \gamma $. Meanwhile, removed individuals lose their immunity at a constant rate $ \nu $. Given these parameters, the basic reproductive ratio is defined as $ R_{0}=\frac{\beta}{\gamma} $. Herd immunity is considered an important concept for an infection when susceptible individuals have enough immunity to prevent the spread of the infection. Thus, let the condition for herd immunity to occur be $ R_{0} S < 1 $.

The evolution of the fractions of the population $ S(t) $, $ I(t) $, and $ R(t) $ can be described as a dynamical system
 \begin{equation}\label{1}
 \begin{split}
 & \frac{dS}{dt}=-\beta SI + \nu R,\\ &
 \frac{dI}{dt}=\beta SI - \gamma I,\\ &
 \frac{dR}{dt}=\gamma I - \nu R,\\
 \end{split}
  \end{equation}
which has two equilibria, firstly, the disease-free equilibrium $ (1,0,0) $ that is stable when $ R_{0} < 1 $ and unstable when $ R_{0} > 1 $, and secondly, the endemic equilibrium
\begin{equation}\label{2}
\left(S,I,R\right) =\left( \dfrac{\gamma}{\beta}, \dfrac{1}{1+\dfrac{\gamma}{\nu}} \left[ 1-\dfrac{\gamma}{\beta}\right] , \dfrac{1}{1+\dfrac{\nu}{\gamma}}\left[ 1-\dfrac{\gamma}{\beta}\right]\right),
\end{equation}
provided that $ R_{0} > 1 $. The endemic equilibrium is stable when present.

An infection can be contained if the disease-free equilibrium is the only possible stable equilibrium. For the COVID-19, $ R_{0} =3.28 $ \cite{9}, which implies that the disease-free equilibrium is found to be unstable. If the endemic equilibrium is reached, it will be stable, the disease will persist, and herd immunity will not be effective. Thus, there is an urgent need to contain the COVID-19 before the endemic equilibrium is reached. The two most important strategies to control the transmission are SD and vaccination.

%\section{The model with SD and vaccination}
The COVID-19 has been found to spread even during the incubation period \cite{8}. The policies of SD reduce the probability of contact with both infective individuals and polluted objects. SD can be integrated in the SIRS model, by assuming that only $ (1-\mu) $ of the susceptible individuals are in contact with the infective individuals, where $ 0\le \mu \le 1 $ is the degree of SD. Here, the dynamical system (\ref{1}) transforms to
\begin{equation}\label{3}
\begin{split}
& \frac{dS}{dt}=-\beta (1-\mu) SI + \nu R,\\ &
\frac{dI}{dt}=\beta (1-\mu) SI - \gamma I,\\ &
\frac{dR}{dt}=\gamma I - \nu R.\\
\end{split}
\end{equation}

Note that the reproductive ratio becomes $ (1-\mu)R_{0} $, which implies that the infectiousness of COVID-19 can be lessened by increasing the degree of SD. Further, it can be reduced to half for $ \mu = 0.5 $. Accordingly, two equilibria are present: the disease-free equilibrium that is stable when $ (1-\mu)R_{0} < 1 $ and the stable endemic equilibrium
\begin{equation}\label{4}
\left( S,I,R\right) =\left( \dfrac{\gamma}{\beta (1-\mu)}, \dfrac{1}{1+\dfrac{\gamma}{\nu}} \left[ 1-\dfrac{\gamma}{\beta (1-\mu)}\right], \dfrac{1}{1+\dfrac{\nu}{\gamma}}\left[ 1-\dfrac{\gamma}{\beta(1-\mu)}\right]\right),
\end{equation}
provided that $ (1-\mu)R_{0} > 1 $. Again, herd immunity is not applicable if the endemic equilibrium has already been reached. To avoid falling in the stable endemic equilibrium, SD must be applied with a degree
\begin{equation}\label{5}
\mu \ge 1-\dfrac{1}{R_{0}}.
\end{equation}
The sufficient SD degree to control the infection can be defined as $ \mu_{0} = 1-\dfrac{1}{R_{0}} $. For COVID-19, $ \mu_{0} = 0.69 $. A degree higher than $ \mu_{0} $ should be maintained for a sufficient time to ensure the robustness of the stability of the disease-free equilibrium, and to not reach the endemic equilibrium.

An interesting limit is when $ \mu = 1 $. From system (\ref{3}), because $ I $ exponentially decays as
\begin{equation}\label{6}
I(t)=I(0)\mathrm{e}^{-\gamma t},
\end{equation}
the spread of the infection is stopped.

\begin{figure}
	
	\begin{center}	
		\includegraphics[width=\textwidth]{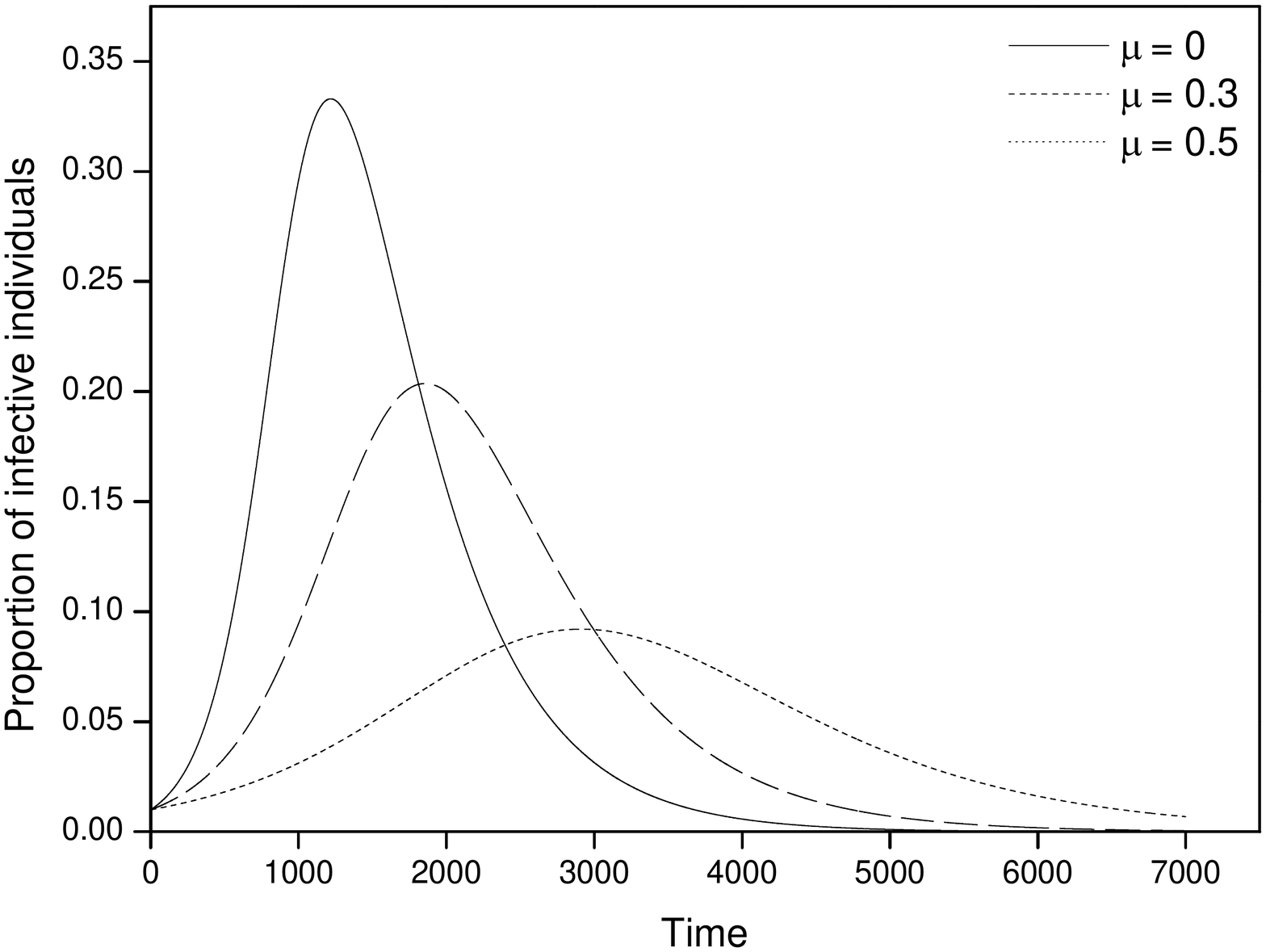}
	\end{center}	
	
	\caption{Numerical simulations for the SIRS model taking into consideration social distancing with $ \beta = 0.65 $, $ \gamma = 0.02 $, $ \nu=0.05 $ and different degrees of social distancing $ \mu=0,\quad 0.3, \quad 0.5 $. The proportion of infective individuals, $ I $ is plotted versus time. For $ \mu=0 $, $ I $ increases exponentially until reaching its peak at $ I_{max}=0.333 $ with a peak time $ t_{Imax}=1218 $ time steps. Then it decreases exponentially until the infection is eradicated. As $ \mu $ increases, the peak of infection is reduced and the peak time is delayed.} \label{F1}
		
\end{figure}

Figure \ref{F1} shows the results of the computations when $ \beta = 0.65 $, $ \gamma = 0.02 $, and $ \nu=0.05 $. In the absence of SD policies ($ \mu=0 $), the proportion of infective individuals $ I $ increases exponentially with time until the peak ($ I_{max}=0.333 $) is reached at time ($ t_{Imax} = 1218 $ time steps). In contrast, because of herd immunity, $ I $ decreases exponentially with time until the spread of the virus is prevented. $ I $ has been determined to respond differently with various degrees of SD applied. For $ \mu=0.3 $, $ I $ increases at a slower rate than when $ \mu=0 $, the peak is reduced by $ 39\% $, and the peak time delayed by $ 53\% $ ($ I_{max}=0.204 $ and $ t_{Imax} = 1865 $ time steps). For $ \mu=0.5 $, $ I_{max}=0.092 $ and $ t_{Imax} = 2911 $ time steps, see Fig. \ref{F1},
when $ \mu $ is increased, the peak is reduced correspondingly (dues to lower herd immunity threshold), and there is more delay in the peak time. These results are important as they offer the opportunity for more studies and more time to develop a vaccine, while simultaneously reducing the patient loads in hospitals and intensive care units.

At the beginning of an epidemic, $ S $ is approximately the whole population, and can be considered as a constant $ S(0) $, then from system (\ref{3})
\begin{equation}\label{7}
I(t)=I(0)\mathrm{e}^{\beta (1-\mu) S(0) t},
\end{equation}
which explains the initial exponential growth of $ I $.

\begin{figure} 	
	
	\begin{center}	
		\includegraphics[width=\textwidth]{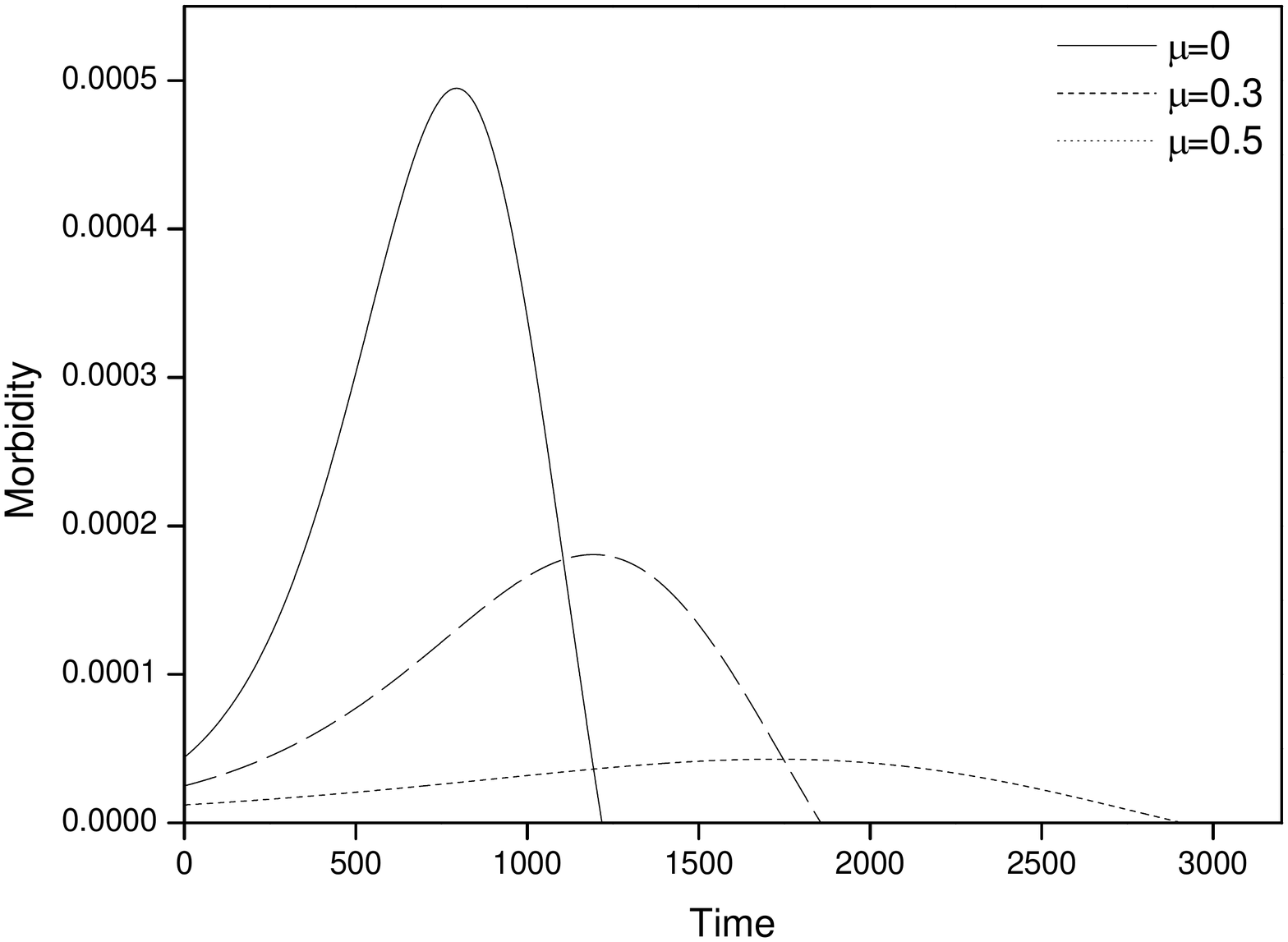}
	\end{center}	
			\caption{Numerical simulations for the morbidity $ m(t) $ with $ \beta = 0.65 $, $ \gamma = 0.02 $, $ \nu=0.05 $ and different degrees of social distancing $ \mu=0,\quad 0.3, \quad 0.5 $. The morbidity and thus mortality decreases as $ \mu $ increases.} \label{F2}
\end{figure}

Both morbidity and mortality are commonly used measures that determine the severity of an infection. The morbidity $ m(t) $ is calculated as
\begin{equation}\label{8}
m(t)=I(t)-I(t-1).
\end{equation}
Figure \ref{F2} shows the results of the numerical simulations for $ m(t) $, with the same values of the parameters of the model as Fig. \ref{F1}. It is clear that increasing $ \mu $ decreases morbidity, and thus mortality associated with COVID-19.

In the presence of a vaccine, if a fraction $ p $ of the susceptible individuals is vaccinated, then only $ 1-p $ of them will be exposed to infection; the reproductive ratio becomes $ (1-p)(1-\mu)R_{0} $. The sufficient vaccination ratio for controlling an infection is modeled by
\begin{equation}\label{9}
p_{0} = 1-\dfrac{1}{(1-\mu)R_{0}}
\end{equation}
which depends on the infectiousness of the disease and the SD degree applied. The ratio $ p_{0} $ decreases rapidly with an increase in $ \mu $, and it approaches zero when $ \mu = \mu_{0} $. This is advantageous as vaccination can be applied to only a small fraction of the population, provided that SD policies are applied.

References \cite{13} and \cite{14} model mistakes by adding coupling terms to system (\ref{3}); these coupling terms represent the individuals' interaction with the environment. Here assume that the degree of community awareness is $ 0 \leqslant \alpha \leqslant 1 $. As a general rule, the least aware community would commit more mistakes. The dynamical system (\ref{3}) is modified to the form
\begin{equation}\label{10}
\begin{split}
& \frac{dS}{dt}=-\beta (1-\mu) SI + \nu R + (1-\alpha) S \left[ I \mathrm{ln}\left( \dfrac{I}{S}\right) + R \mathrm{ln}\left( \dfrac{R}{S}\right)\right] ,\\ &
\frac{dI}{dt}=\beta (1-\mu) SI - \gamma I + (1-\alpha) I \left[ S \mathrm{ln}\left( \dfrac{S}{I}\right) + R \mathrm{ln}\left( \dfrac{R}{I}\right)\right],\\ &
\frac{dR}{dt}=\gamma I - \nu R + (1-\alpha) R \left[ S \mathrm{ln}\left( \dfrac{S}{R}\right) + I \mathrm{ln}\left( \dfrac{I}{R}\right)\right].\\
\end{split}
\end{equation}

\begin{figure}
	
	\begin{center}	
		\includegraphics[width=\textwidth]{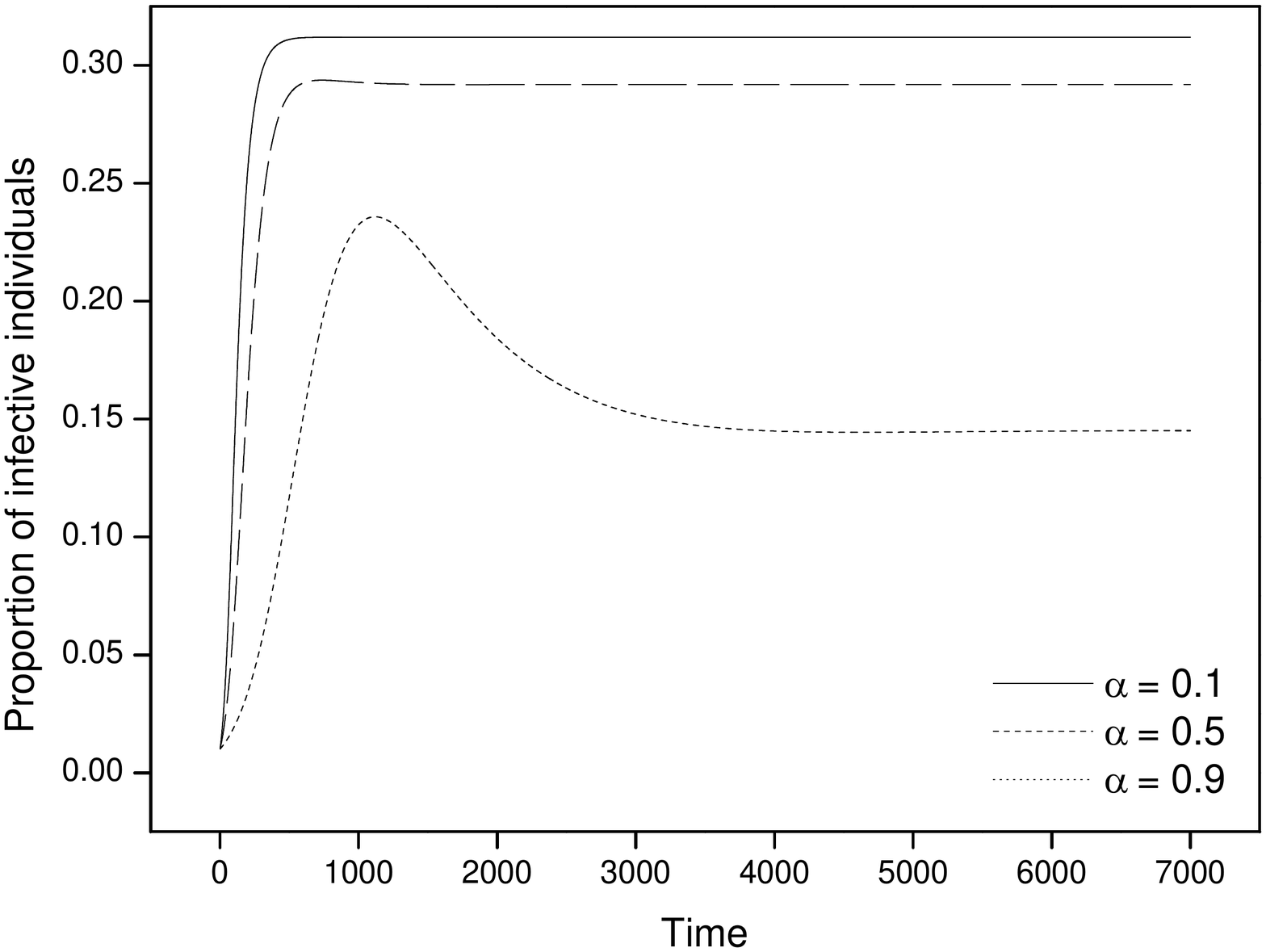}
	\end{center}	
	
	\caption{Numerical simulations for the model taking into consideration both social distancing and community awareness. The values of the parameters are set as $ \mu=0.3 $, $ \beta = 0.65 $, $ \gamma = 0.02 $, $ \nu=0.05 $. Different levels of community awareness $ \alpha=0.9,\quad 0.5, \quad 0.1 $ are studied. As the level of community awareness is lowered, the peak of infection is heightened, and reaches an endemic phase.} \label{F3}
	
\end{figure}

Figure \ref{F3} presents a numerical investigation for system (\ref{10}) when $ \mu = 0.3 $, $ \beta = 0.65 $, $ \gamma = 0.02 $, $ \nu=0.05 $, and a varied $ \alpha $. Note the advantages of SD are lost once the degree of community awareness is slightly lowered to $ 0.9 $; moreover, the peak increases to $ 0.236 $, and herd immunity is destroyed, allowing the infection to preset. Further, the infection is expected to get worse when the community awareness drops to $ \alpha = 0.5 $ and then to $ \alpha = 0.1 $ (see Fig. \ref{F3}). This implies that the infection will grow through the population, which confirms the important role of community awareness development in reducing COVID-19 transmission.

\begin{figure}
	
	\begin{center}	
		\includegraphics[width=\textwidth]{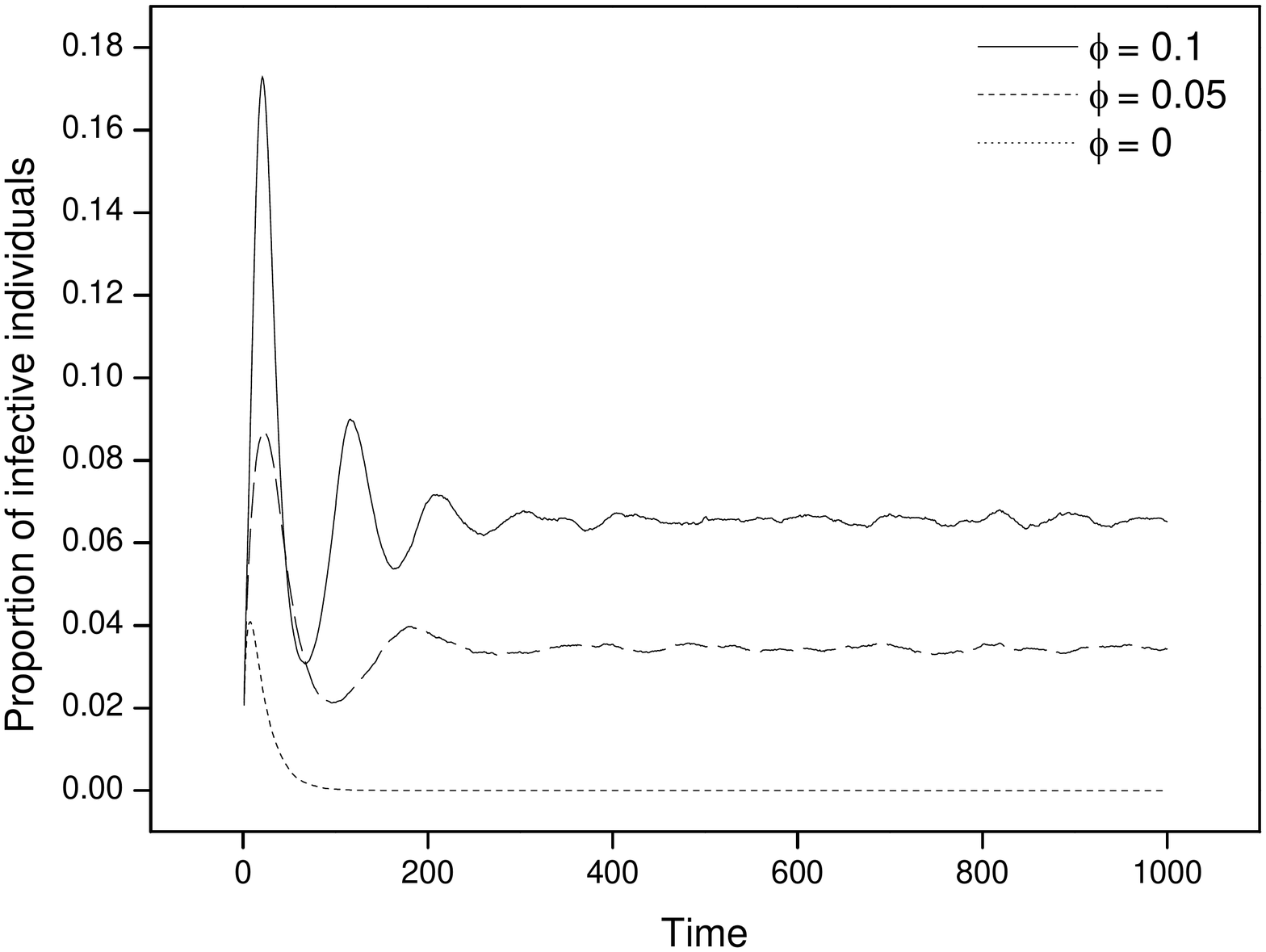}
	\end{center}	
	
	\caption{Simulations of the model on small-world networks. The values of the parameters are set as $ \beta = 0.65 $, $ \gamma = 0.02 $, $ \nu=0.05 $. Different values of the shortcuts densities $ \phi=0,\quad 0.05, \quad 0.1 $ are studied. As the value of $ \phi $ is increased, the infection becomes endemic. Social distancing policies are expected to lower the $ \phi $ value preventing reaching the endemic phase.} \label{F4}
	
\end{figure}

Finally, the spatial aspect is included by simulating the model on small-world networks. In our simulations, a connected one-dimensional lattice of size $ 10^{6} $, with periodic boundary conditions is assumed. With probability $ \phi $, some randomly chosen vertices are connected by some shortcuts to randomly selected non-nearest neighboring vertices. Every vertex interacts with its nearest neighbors and the distant neighbor, if present. It is clear that SD practices reduces the chance for shortcuts. The numerical simulations are performed using the same values of the model (system (\ref{1})) and different values of $ \phi $. For $ \phi = 0 $, the behavior of $ I $ is approximately similar to that of Fig. \ref{F1} for $ \mu=0 $. As $ \phi $ increases, the infection becomes endemic see Fig. \ref{F4}. Of course, applying SD policies reduces $ \phi $, therefore do not allow reaching the endemic phase.

\section{Nonhomogeneous population}\label{s3}

The nonlocal interaction is widely observed in the epidemic spreading, especially in the recent COVID-19. Here we use the Lajmanovich-Yorke model \cite{4} which combines between both local and nonlocal epidemic spreading. The population is divided into $ n $ patches. Let $ I_{i} $ be the fraction of infective individuals in the $ i $-th patch, $ \beta_{ji} $ be the infection rate within the $ i $-th patch due to an infection from the $ j $-th patch, and $ \gamma_{i} $ be the recovery rate of the $ i $-th patch individuals. Then the model is given as follows
\begin{equation}\label{11}
\frac{dI_{i}}{dt}=\sum_{j=1}^{n} \beta_{ji} I_{j}-\sum_{j=1}^{n} \beta_{ji} I_{i} I_{j}-\gamma_{i}I_{i}, i=1,2,\dots,n.
\end{equation}
Let $ \mathbf{I}=\left(I_{1},I_{2},\dots,I_{n} \right)^{T} $, and $ A=(a_{ij}) $, such that $ a_{ij}= \beta_{ji}, i\neq j$; and $ a_{ii}= \beta_{ii}-\gamma_{i}$, then the system (\ref{11}) can be written as follows
\begin{equation}\label{12}
\frac{d\mathbf{I}}{dt}=A \mathbf{I}-\sum_{j=1}^{n} \beta_{ji} I_{i} I_{j}.
\end{equation}
Let $ \lambda_{max} = \max_{1\leq i\leq n} \mathrm{Re}(\lambda_{i}) $, where $ \lambda_{i}, i=1,2,\dots,n $ are the eigenvalues of $ A $. According to Ref. \cite{4}, there are two possibilities: either the disease-free equilibrium $ (I_{i}=0, i=1,2,\dots,n) $ is asymptotically stable if $ \lambda_{max} \leq 0 $, or there exists an asymptotically stable endemic equilibrium when $ \lambda_{max} > 0 $.

When $ n=1 $, the system is reduced to the homogeneous population situation. The case $ n=2 $ is worthy of investigation because it represents the situation when the population is divided into two subpopulations: inside and outside a certain region. For simplicity, we can assume $ \beta_{11}=\beta_{22} =\beta $, $ \beta_{12}=\beta_{21}=\acute{\beta} $, and $ \gamma_{1}=\gamma_{2}=\gamma $. The system (\ref{11}) becomes
\begin{equation}\label{13}
\begin{split}
\frac{dI_{1}}{dt}= \left(\beta-\gamma\right) I_{1}+ \acute{\beta} I_{2}- I_{1}\left(\beta I_{1}+\acute{\beta} I_{2}\right),\\
\frac{dI_{2}}{dt}= \left(\beta-\gamma\right) I_{2}+ \acute{\beta} I_{1}- I_{2}\left(\beta I_{2}+\acute{\beta} I_{1}\right).
\end{split}
\end{equation}
The conditions for the asymptotic stability of the disease-free equilibrium become
\begin{equation}\label{13}
\beta-\gamma<0,\quad \left(\beta-\gamma\right)^{2}-\acute{\beta}^{2}\ge 0.
\end{equation}
Then
\begin{equation}\label{14}
\frac{\beta+\acute{\beta}}{\gamma} \le 1.
\end{equation}
Equation (\ref{14}) can be written as
\begin{equation}\label{15}
R_{0}+\acute{R}_{0} \le 1,
\end{equation}
where $ R_{0} $ and $ \acute{R}_{0} $ are local and nonlocal reproductive ratios, respectively.

\begin{figure}
	
	\begin{center}	
		\includegraphics[width=\textwidth]{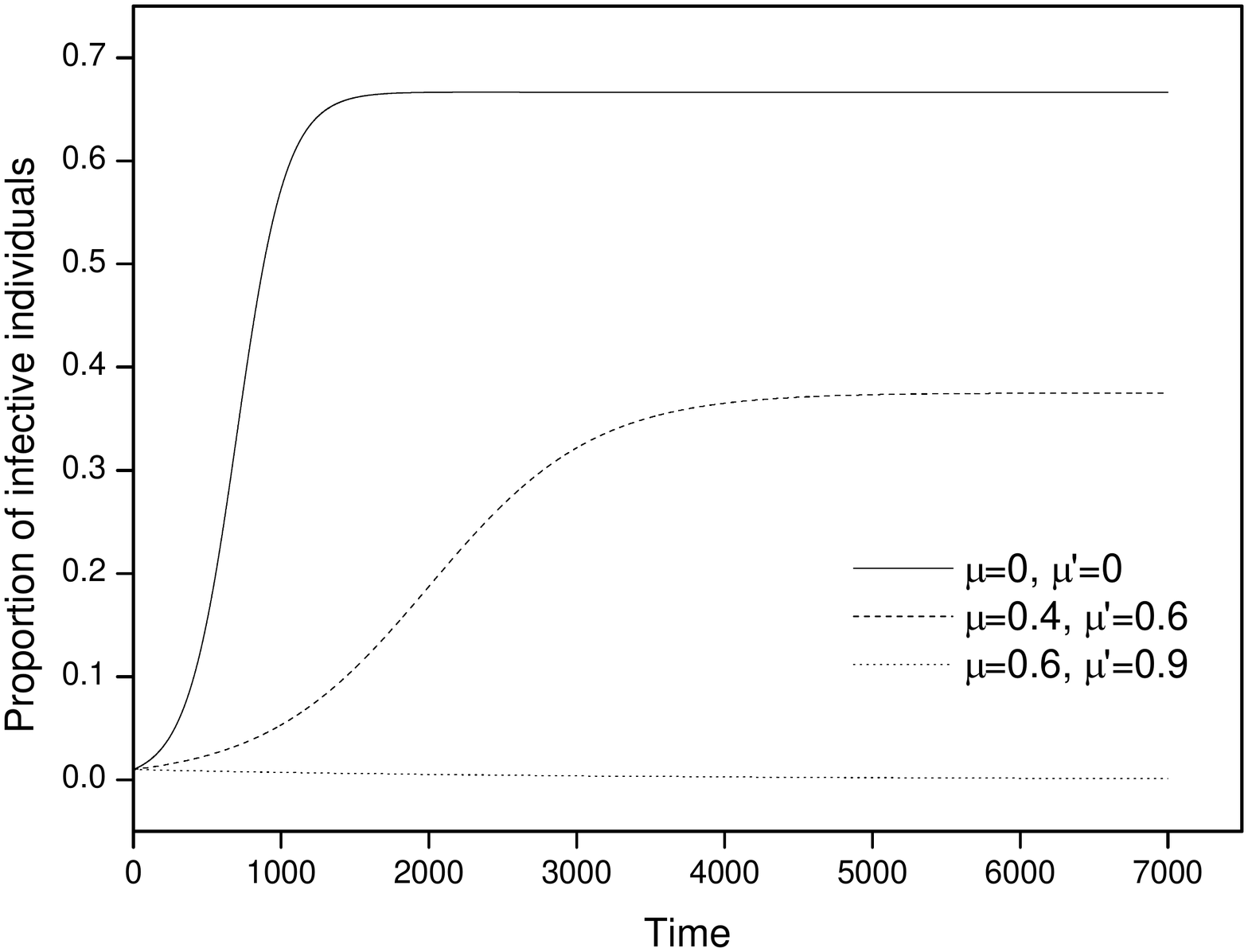}
	\end{center}	
	\caption{Investigation of the effect of both internal and external SD policies $ \mu $ and $ \acute{\mu} $. The other variables are setted as follows: $ \beta = 0.6 $, $ \acute{\beta} = 0.3 $ and $ \gamma = 0.3 $. Obviously, the proportion of infective individuals in the endemic phase decreases with increasing the SD degrees $ \mu $ and $ \acute{\mu} $. The disease-free equilibrium can be obtained at suitable values for $ \mu $ and $ \acute{\mu} $.} \label{F5}
	
\end{figure}

By applying a SD program with degrees, $ \mu $ locally and $ \acute{\mu} $ nonlocally, the condition for the asymptotic stability of the disease-free equilibrium become
\begin{equation}\label{16}
(1-\mu)R_{0}+(1-\acute{\mu})\acute{R}_{0} \le 1.
\end{equation}
Then
\begin{equation}\label{17}
\mu \ge 1-\frac{1}{R_{0}}+\left(1-\acute{\mu}\right)\frac{\acute{R}_{0}}{R_{0}}.
\end{equation}
Therefore, applying an external SD program with an appropriate degree reduces the required degree of the internal SD program to prevent approaching the endemic equilibrium. This is beneficial from a domestic economy point of view. The effect of both internal and external SD policies is investigated in Fig. \ref{F5}. The variables are setted as follows: $ \beta = 0.6 $, $ \acute{\beta} = 0.3 $ and $ \gamma = 0.3 $. It is clear that the proportion of infective individuals in the endemic phase is lowered by increasing the SD degrees $ \mu $ and $ \acute{\mu} $. Also, the disease-free equilibrium can be obtained at suitable values for $ \mu $ and $ \acute{\mu} $.

\section{Conclusion}\label{s4}
In the absence of a vaccine, an appropriate degree of SD and a high level of community awareness can help in controlling the spread of COVID-19. SD policies in particular can reduce the peak of infection and delay the peak time leading to offer ample time for the development of a vaccine, and prevent hospitals from reaching their maximum patient capacity. Moreover, SD policies can reduce the effects of non-local interactions that can turn the infection to be endemic. With a vaccine, SD practices can reduce the vaccination rate required to eradicate infection. Subsequently, vaccine priority can be given to the health-care providers and those in direct contact with infective individuals. Long-term SD practice is crucial to avoid a relapse, as immediately loosening SD rules may provide an avenue of recurrence of the infection. Community awareness is also important in the transmission process; low community awareness destroys the herd immunity, which spreads the infection endemically. Appropriately implementing an external SD program reduces the degree required for the internal SD program to contain the infection, which has a positive impact on the local economy.

There are some probable risks. Firstly, COVID-19 can become endemic especially in the poorest regions around the world. Secondly, COVID-19 will return in some regions; this case often refers to the second wave of infection. Thirdly, the possibility of creation of new strains, which can be slowed down as the number of infected individuals decreases. When such risks happen, the virus gets retransmitted to the recovered areas. Thus, it is important that poor countries should be assisted in controlling the transmission of the virus. Appropriate levels of long-term external and internal SD practices are also strongly recommended.

Some SD practices can benefit in other issues. For example, COVID-19, though not sexually transmitted, can be passed easily through sexual activities with an infective person. Thus, SD practices only allow spouses living in the same household to perform sexual activities. This will help powerfully in reducing the sexually transmitted infections which have a profound impact on human health.

%\section*{Acknowledgements}
%The author is very grateful to the reviewers for the helpful suggestions.

%\section*{Declaration of Competing Interest}
%The author declares that he has no known competing financial interests or personal relationships that could have appeared to influence the work %reported in this paper.

\newpage

\newpage

\newpage

\newpage

\newpage

\begin{thebibliography}{99}
\bibitem{1} H.W. Hethcote, The mathematics of infectious diseases, SIAM Rev. 42 (2000) 599-653.
\bibitem{2} R.W. West, J.R. Thompson, Models for the simple epidemic, Math. Biosci. 141, (1997) 29-39.
\bibitem{3} E. Ahmed, A.S. Hegazi, A.S. Elgazzar, An epidemic model on small-world networks and ring vaccination, Int. J. Mod. Phys. C 13 (2002) 189-198.
\bibitem{4} A. Lajmanovich, J.A. Yorke, A deterministic model for Gonorrhea in a nonhomogeneous population, Math. Biosci. 28 (1976) 221-236.
\bibitem{5} World Health Organization. Novel coronavirus (2019-nCoV) situation reports, \url{https://www.who.int/emergencies/diseases/novel-coronavirus-2019/situation-reports}, 2020 (accessed 15 December 2020).
\bibitem{6} C. Wang, P.W. Horby, F.G. Hayden, G.F. Gao, A novel coronavirus outbreak of global health concern, The Lancet, 395 (2020) 470-473.
\bibitem{7} Y. Wang, Y. Wang, Y. Chen, Q. Qin, Unique epidemiological and clinical features of the emerging 2019 novel coronavirus pneumonia (covid-19) implicate special control measures, J. Med. Virol. 92 (2020) 568-576.
\bibitem{8} P. Li, J.B. Fu, K.F. Li, J.N. Liu, H.L. Wang, L.J. Liu, Y. Chen, Y.L. Zhang, S.L. Liu, A. Tang, Z.D. Tong, J.B. Yan, Transmission of COVID-19 in the terminal stages of the incubation period: A familial cluster, Int. J. Infect. Dis. 96 (2020) 452-453.
\bibitem{9} Y. Liu, A.A. Gayle, A. Wilder-Smith, J. Rockl\"{o}v, The reproductive number of COVID-19 is higher compared to SARS coronavirus, J. Travel Med. 27 (2020) 1-4.
\bibitem{10} J.M. Crick, D. Crick, Coopetition and COVID-19: Collaborative business-to-business marketing strategies in a pandemic crisis, Ind. Market. Manag. 88 (2020) 206-213.
\bibitem{11} A.S. Elgazzar, Coopetition in quantum prisoner's dilemma with different dilemma strength and COVID-19, in preparation (2020).
\bibitem{12} S.	Camazine, J.L. Deneubourg, N.R. Franks, J. Sneyd, E. Bonabeau, G.Theraula, Self-Organization in Biological Systems", Princeton University Press, Princeton, 2001.
\bibitem{13} Y. Sato, J.P. Crutchfield, Coupled replicator equations for the dynamics of learning in multiagent systems, Phys. Rev. E 67 (2003) 015206(R).
\bibitem{14} E. Ahmed, A.S. Hegazi, A.S. Elgazzar, Sato-Crutchfield formulation for some evolutionary games, Int. J. Mod. Phys. C 14 (2003) 963-971.
\bibitem{15} B. Ivorra, M.R. Ferr\'{a}ndez, M. Vela-P\'{e}rez, A.M. Ramos, Mathematical modeling of the spread of the coronavirus disease 2019 (COVID-19) taking into account the undetected infections. The case of China, Commun. Nonlinear Sci. Numer. Simulat. 88 (2020) 105303.
\bibitem{16} K. Roosa, Y. Lee, R. Luo, A. Kirpich, R. Rothenberg, J. M. Hyman, P. Yan, G. Chowell, Real-time forecasts of the COVID-19 epidemic in china from February 5th to February 24th, 2020, Infect. Dis. Model. 5 (2020) 256-263.
\bibitem{17} A.J. Kucharski, T.W. Russell, C. Diamond, Y. Liu, J. Edmunds, S. Funk, R.M. Eggo, F. Sun, M. Jit, J.D. Munday, J.D. Davies, Early dynamics of transmission and control of COVID-19: a mathematical modelling study, Lancet Infect. Dis. 20 (2020) 553-558.
\bibitem{18} A.W.D.Edridge,J.M. Kaczorowska, A.C. Hoste, M. Bakker, M.Klein, M.F.Jebbink, A. Matser, C. Kinsella, P. Rueda, M. Prins, P. Sastre, M. Deijs, L. van der Hoek, Human coronavirus reinfection dynamics: lessons for SARS‐CoV‐2, doi: \url{https://doi.org/10.1101/2020.05.11.20086439}, (2020).
\bibitem{19}R. Pastor-Satorras, C. Castellano, P.V. Mieghem, A. Vespignani, Epidemic processes in complex networks, Rev. Mod. Phys. 87 (2015) 925-979.
\bibitem{20} D.J. Watts, S.H. Strogatz, Collective dynamics of small-world networks, Nature 392 (1998) 440-442.
\bibitem{21} S. R\"{u}diger, A. Plietzsch, F. Sagu\'{e}s, I.M. Sokolov, and J. Kurths, Epidemics with mutating infectivity on small-world networks, Sci. Rep. 10 (2020) 5919.

\end{thebibliography}
\end{document}